\documentstyle[12pt,aasms4]{article}
\slugcomment{Submitted to: The Astrophysical Journal (Letters) }
\lefthead{Che }
\righthead{Constraints on SN GRBs}

\begin{document}

\title{Constraints on Jets and Luminosity Function of Gamma-ray Bursts Associated with Supernovae}

\author{Haihong Che}
\affil{Department of Physics, Michigan Technological University, 1400 Townsend Dr., Houghton, MI, 49931, hche@mtu.edu}

\begin{abstract}
If Gamma-ray Bursts (GRBs) are generally associated with supernovae like 1998bw, a relatively wide intrinsic luminosity function is implied, which indicates the existence of a large amount of undetected dim bursts, and a much higher event-rate than is often assumed.   If it is assumed that the intrinsic luminosity function of GRBs is a power-law: ${\phi (L)} \propto {L^{-\beta}}$ ($\beta > 0$, $L_{min} \leq L \leq L_{max}$), data from the BATSE 4B catalog can be used to constrain slope index $\beta$ and the dynamic range width  Log ${L_{max}\over L_{min}}$. Using a K-S test comparison with the observational Log $N$ - Log $P$, we find constraints on the GRB fireball model, GRB jets, and the possible GRB contribution to cosmic gamma-ray background.  We find the acceptable dynamic range for $10^2<$$L_{max}/L_{min}$$ < 10^7$.Our results show that  jet model is more likely to be related more highly energetic explosion than fireball model.Our studies also show that the luminosity function provided by a purely special relativistic effect on a jet is outside of the K-S test acceptable range.  Were intrinsic beaming to confine the jet to $\theta_{max} < 1/\gamma$ however, the effects of relativistic beaming would not dominate.

\end{abstract}

\keywords{gamma ray bursts - stars: formation - cosmology: observations - supernovae: general}

\section{ Introduction }
Since the suggested association between a Gamma-Ray Burst (GRB) and Supernova 1998bw (Galama et al. 1998), the possible general connection between GRBs and SNe has been discussed (Cen 1998, Wang et al. 1998).   In contrast, Kippen et al. (1998) found no evidence for a GRB - SNe connection, and Grazianni, Lamb \& Marion (1998) even conclude that not all SN Ib,c/II can be associated with GRBs, although a fraction might be. On the other hand, Wang et al. (1998) found seven candidate SN-GRB correlations while Bloom et al. (1998) estimated these SN-GRB isotropic explosion energy range from $10^{44} - 10^{48}$ erg/s, and suggested that a subclass of GRBs is associated with supernova Ib,c/II. Jets have been introduced to solve two problems raised in the GRB-SN united picture since 1) the great difference between the BATSE-inferred GRB rate with a narrow luminosity function ($\sim 10^{-8}$ h$^{3}$ Mpc$^{-3}$ yr$^{-1}$) (Kommers et al. 1999; Schmidit 1999) and the supernova rate (
$\sim 10^{-4}$ h$^{3}$ Mpc$^{-3}$ yr$^{-1}$), and 2) the great difference between the explosion energy of GRBs and supernovae.  With a jet the total energy necessary to produce the observational fluxes will be reduced by a factor $1/\gamma ^{2}$ ($\gamma$ is the Lorentz factor of a relativistic jet), and the total rate will be greater by a factor $\gamma^{2}$ (Cen 1998; Wang et al. 1998).  Were a wide GRB luminosity function or jet to operate, BATSE-detected GRBs would only be a small fraction of all GRBs that occur.

In this Letter, we compare GRB rates derived from supernova rates and density evolution to that of the measured rates from the Fourth BATSE Catalog of GRBs to investigate possible constraints on the GRB luminosity function and jet parameters.
We also discuss two other related interesting questions: 1) Can the special relativistic effects of a jet provide by itself an acceptable luminosity function for GRBs? 2) Can the large amount of undetected bursts contribute to the observational gamma-ray background near 100 KeV for an acceptable GRB luminosity function? The former provides an alternative way to test if the gamma-ray burst emission is isotropic or random in the jet's comoving frame.

\section{Model}
If supernovae are the mysterious sources of GRBs, then the GRB "equivalent isotropic" explosion energy could be within a relatively large range: from $\sim 10^{54}$ erg/s, deduced for GRB 990123 (Kulkarni et al 1999) to $\sim 10^{44}$ erg/s estimated from SN Ib/c - GRBs associations (Wang \& Wheeler 1998; Bloom et al. 1998).  We assume the unknown intrinsic luminosity distribution follows a power-law:
\begin{equation}
\phi (L) = A  \ {({L \over L_{max}})}^{-\beta}, ~~~~~~{L_{min} \leq L \leq L_{max}},~~~~~~~ {\beta > 0},
\end{equation}
where $A$ is the normalization factor. We fix $L_{max}=10^{53}$ erg/s. We find that varying $L_{max}$ from $10^{53}$ to $10^{55}$ erg/s affected our results only slightly; $L_{min}$ was allowed to vary in the range $10^{44}$ - $L_{max}$.   Observed spectra of GRBs can be approximated as a power-law: $dL(E)/dE \propto E^{- \alpha}$ with spectral index $\alpha = 1.1$ (Mallozzi et al. 1996).  For source density, we adopt the broken power-law function we have used before as a rough description for SN Ib/c evolution (Che et al. 1999):

$n(z)={n_{0} \left( 1+z\over 1+z_{0}\right)^\eta } \cases{\eta =\eta_{1} >0, z<z_{0}\cr
                                                            \eta=\eta_{2} <0, z>z_{0}\cr}$

\noindent
where $n_{0}$ is the comoving GRB density at redshift $z=z_{0}$ per Mpc$^3$ yr$^{-1}$. We set  $z_{0}=1$,  $\eta_{1} = 4$ (Lilly 1996), and $\eta_{2} = 0$ (Madau 1998, Sadat 1998).

We will assume a universe where the calculations are particularly simple: a flat Friedmann universe with cosmological constant $\Lambda=0$.  We do not expect the results would differ significantly in universes with a significant $\Lambda$.  In our asssumed universe, the number of observed gamma-ray bursts peak flux brighter than P is:

\begin{equation}
N(>P) = { \int_{L_{min}}^{L_{max}} }
              {\phi (L) dL}
              {\int_{0}^{z_{max}(L,P)}}
              {4 \pi n(z)\over 1+z}
              {r^2(z)}
              {dr(z) \over dz} dz
\end{equation}
The observed peak flux of a GRB at redshift $z$ with intrinsic luminosity $L$ is:
\begin{equation}
P(L,z) ={L (1+z)^{- \alpha} /4\pi r^2(z)}
\end{equation}
where we fixed the spectral index to be $\alpha=1.1$ (Mallozzi et al. 1996), and $P_{min} \sim 2 \times 10^{-7}$ erg cm$^{-2}$ s$^{-1}$ $\sim 1$ photon cm$^{-2}$ s$^{-1}$.
\begin{equation}
L= {\int_{50}^{300} {dL(E)\over dE}dE}
\end{equation}
where $E$ is measured in KeV and $r(z)$ is proper motion distance. The total number of bursts we observe every year is
$N_{total}(>P_{min})$ as derived from Eqn. 2.

\section{Results}

\subsection{Luminosity Function Constraints from Log $N$ - Log $P$}
We selected the 939 bursts with peak flux greater than $P_{min}=2 \times 10^{-7} erg/cm^2 s$ from the Fourth BATSE GRB catalog (BATSE 4B) to get the measured Log $N$ - Log $P$ relation that will be used for comparison.  We compared this Log $N$ - Log $P$ to GRBs following the evolving supernova rate, finding the best fits to the luminosity function range parameter $L_{min}$ and slope parameter $\beta$.  The rate was normalized by $N_{total}$ as given in Eq. 5.  We searched the acceptable range of $L_{min}$ and $\beta$ with the K-S confidence level $1 \%$.  The luminosity function width was allowed to vary from $1 < L_{max}/L_{min} < 10^9$, while the luminosity function slope was allowed to vary from 
$0 < \beta < 6$.  The search result is shown in Figure 1: the acceptable range of luminosity function widths were $L_{max}/L_{min} > 100$, while the acceptable rage for the luminosity function slope was $1.8 < \beta < 2.6$.

\subsection{Constraints on GRB Fireballs and Jets}
Setting the $N_{total} = 800$ yr$^{-1}$, which is the BATSE yearly rate corrected for the duty cycle (Meegan et al. 1992), and constraining Log ${L_{max}\over L_{min}}$ and $\beta$ from equation (5), we can find lines of constant rate ($n_0$) in the $\beta$ - Log ${L_{max}\over L_{min}}$ frame.  If we further suppose that GRBs are connected with supernovae, then for the fireball model, the GRB event rate should follow the supernovae rate $\sim 10^{-4}$ h$^{3}$ Mpc$^{-3}$ yr$^{-1}$.  For a jet with a Lorentz factor $\gamma = 100$, however, the GRB rate will decrease to $\sim 10^{-8}$ h$^{3}$ Mpc$^{-3}$ yr$^{-1}$. 

Two lines of constant GRB rate are plotted in Figure 1 and labeled as lines II and III, representing isotropically emitting fireball GRBs and GRBs beamed into a $\gamma=100$ jets respectively.  To the left of these two lines, GRBs will have a higher event rate than detected by BATSE.  For the fireball model, line II traverses the K-S test acceptable region at around $L_{min} \sim  10^{46}$ erg/s, which is enough to produce a gamma-ray burst by shock, but is quite close to the hypothesized theoretical minimum (Meszaros 1999).  Therefore, GRBs occurring at the supernova rate are not in conflict with the isotropic fireball model.  

Line III shows that the minimum isotropic explosion energy of GRBs from a jet is $\sim 10^{49}$ erg/s. From lines II and III, it is not possible to discern whether a jet is preferred from isotropic emission, since both cross the allowed K-S fit region.  Most GRBs in the jet model, however, expend more energy than those in an isotropic fireball model, possibly indicating that jet models are more likely to be related to massive stars (Woosley 1998, Sari 1999) or hypernovae (Paczynski, 1998; Iwamoto et al. 1998; Woosley et al. 1998; Wheeler et al. 1999).

\subsection{Possible Contribution to Cosmic Gamma-ray Background}
As the luminosity function width $L_{max}/L_{min}$ increases, the fraction of undetected bursts become larger, which leads to more GRB photons contributing to the diffuse gamma-ray background near 100 KeV.  Quantitatively, this can be written as :
\begin{equation}
I(E) = {\int_{0}^{\infty}}
{{n(z) \bar{L} (E(1+z))} \over {4 \pi (1+z)^2}}
{dr(z) \over dz} dz ,
\end {equation}
where $\bar L(E(1+z))$ is the average luminosity of individual sources, and $I(E)$ represents the observational gamma-ray background intensity.  We can deduce the required $n_{0}$ from this equation and estimate the required minimum GRB rate from the observation of cosmic ray background (Watanabe et al 1998, Comastri 1998).  This result is shown by line I in Figure 1.
Line I lies outside of K-S test region which indicates that GRBs cannot dominate the diffuse cosmic gamma-ray background.

\subsection{Beaming-Induced Luminosity Function}
A relativistic jet or fireball has been considered as a way to solve the enormous energy release of GRBs (Paczynski \& Xu 1994; Rees \& Meszaros 1994; Sari \& Piran 1997; Pilla \& Loeb 1998). Though the models of shocks (internal or external) for producing a gamma-ray burst usually assume a magnetic field and relativistic particle acceleration (Meszaros et al 1994; Panaitescu et al 1999), the gamma-ray burst radiation should be confined within the beam. The simple assumption that the GRB radiation direction is random in the rest frame of the shocks cannot be excluded absolutely. In this section, we test this assumption by checking if the GRB luminosity function can be induced by a purely special relativistic effect.

We assume that the angle between the line of sight and beaming axis is $\theta$, the velocity of beaming is $v$ and $L_{in}$ is the intrinsic luminosity of GRB in the beaming comoving frame, then the observational flux come from a source at redshift z is (Blandford \& Konigl 1979):
\begin {equation}
P(z,\theta)= {{\Gamma }^{2+\alpha}}
                    {L_{in}\over {4\pi (1+z)^2 r(z)^2}}
\end{equation}
\begin{equation}
\Gamma ={1 \over \gamma ({1-{v \over c} \ {\rm cos} \theta})}
\end{equation}
where $\gamma = 1/\sqrt{1-v^2/c^2}$, the Lorentz factor. We assume the beaming-induced ${(L_{in})}_{min}={\Gamma}^{2+\alpha}_{min} L_{in}$, here ${\Gamma}_{min}$ corresponds to the biggest viewing angle $ {\theta}_{max}$ we can see a burst. ${(L_{in})}_{max}=L_{in} \Gamma^{2+\alpha} (\theta=0)$, seen by the observer within the beam, ${L \over L_{in}} = {\Gamma}^{2+\alpha}$. The induced luminosity function $\phi ({L \over L_{in}})$ is:
\begin{equation}
\phi ({L \over {L_{in}}}) \ d{L \over L_{in}} = 
        {\phi (cos \theta)} {{d {\rm cos} \theta \over 
          d{L \over L_{in} }}}
       {d{L \over L_{in}}}
\end{equation}
\begin{equation}
\phi ({L \over {L_{in}}}) = 
          {{(L/L_{in})}^{- ({3+\alpha\over 2+\alpha})} 
          \over {(2+\alpha) \gamma {v \over c }}}
\end{equation}
With $\alpha=1.1 \pm 0.3$, the induced luminosity function index is constrained to be $1.3 < \beta < 1.36$, far below the required index for Log $N$ - Log $P$. We indicate this allowed region in Figure 1.  We note that this allowed region does not intersect the allowed region created by the necessity of fit to the measured BATSE Log $N$ - Log $P$.  This contradiction implies that the GRB luminosity function is not created by purely special relativistic bulk motion.  

Perhaps the real luminosity function of GRBs has only a component due to beaming.  If so, the beaming induced luminosity function cannot dominate but modulate the intrinsically wide luminosity function.  For this to be true, the intrinsic dynamic range induced by beaming must be quite modest.  From inspection of Figure 1, we see that it if beaming creates only $L_{max}/L_{min} < 100$, it will not dominate the wider intrinsic luminosity function.  From Eqn. 7, we see that if the gamma-ray burst is beamed into an angle of $\theta_{max} \le 1/\gamma$, then the extra amount of beaming generated by relativistic motion will generate an $L_{max}/L_{min} < 100$ for $0 < \theta < \theta_{max}$, and so could be hidden by the intrinsic luminosity function of GRBs. This result can be
tested by detecting the off-axis x-ray emissions from beamed GRBs, caused by  purely special relativistic effects, proposed by Woods and Loeb (1999).

\section { Discussion and Conclusions }
Up to now, the sources for GRBs remain mysterious.  Though some observations suggest that GRBs are associated with supernovae type Ib,c/II, we need more information to confirm this kind of connection. However, the nature of the intrinsic luminosity function is a more general problem. In this letter, we have studied the luminosity function from various aspects and given a clearer sketch on how the luminosity function dynamic range is constrained by the gamma ray burst rate, and what contradictions exist between the observationally favored luminosity and a beaming-induced luminosity function. Our results are based on the assumption of connection between GRBs and SNe.

Our results show 1) if a model for GRBs is considered for the consistence with GRB rate, we cannot ignore the possible intrinsic dynamic range of luminosity function to the rate. 2) If we consider the source density evolution of GRBs, the required width of luminosity function, Log ${L_{max}\over L_{min}}$ cannot be less than 2.  3) The required minimum explosion energy for GRBs from the relative GRB rate, the limits on the width of the GRB luminosity function, and the exclusion of special relativistic motion as the cause of the luminosity function may turn out to be helpful to those studying the GRB radiation and progenitor mechanisms.

This research was supported by grants from NASA and the NSF.  I would like to thank Robert Nemiroff for helpful discussions and a critical reading of the manuscript. I also thank Peter Meszaros for his valurable discussion and comments.

\clearpage

\figcaption[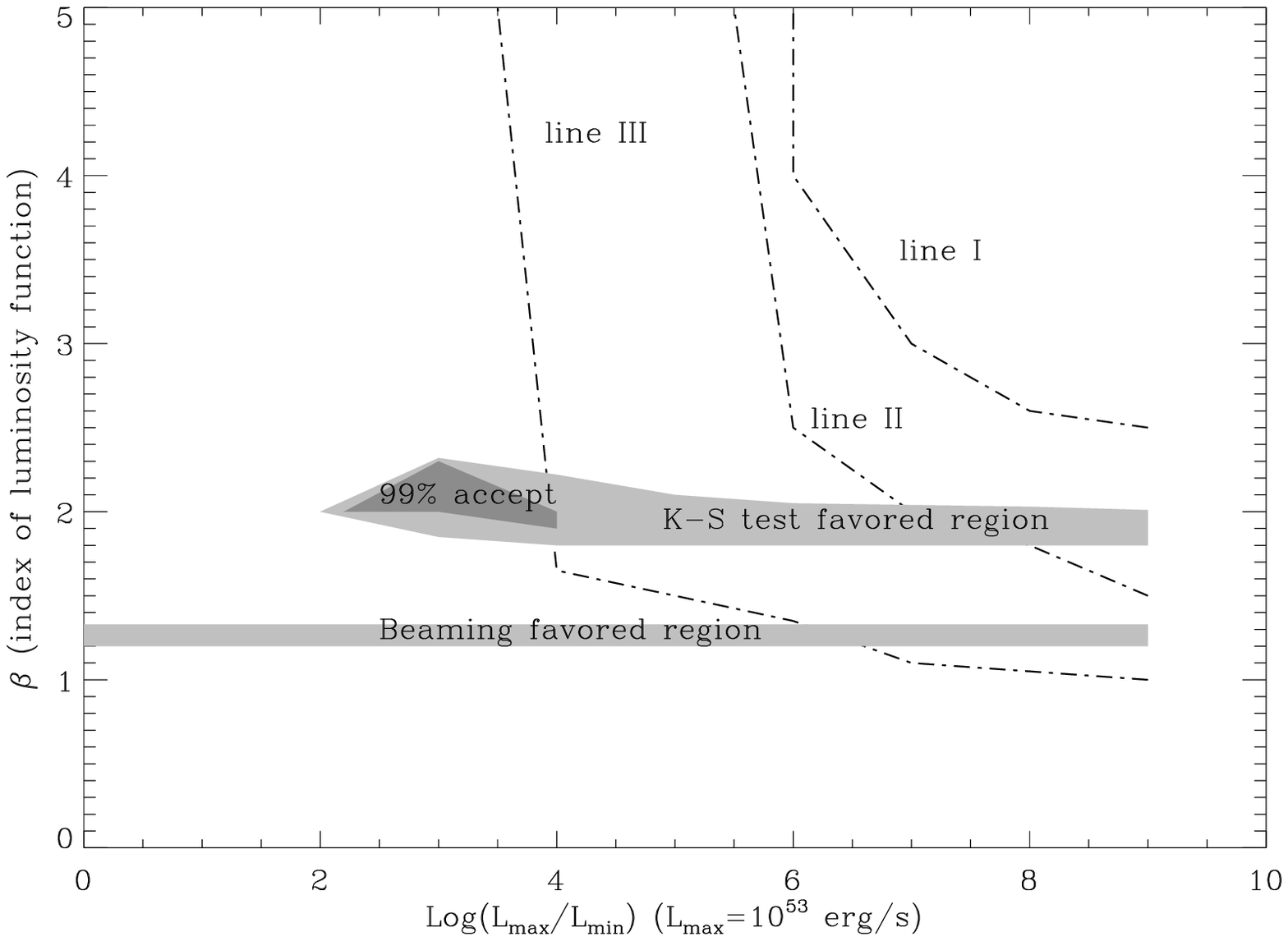]{The shaded area indicates the allowed region of beaming induced GRB luminosity functions to the BATSE Log $N$ - Log $P$.  The boundary of the shaded region corresponds to a K-S confidence level of $1\%$.  Line I
satisfies the relation between the index of Power-Law Luminosity function \& the width $Log {L_{max}\over L_{min}}$ required by the rate producing a comparable observational diffuse gamma-ray background between 50-300 KeV. To the right side of this
line, the corresponding GRB rate will produce too strong a gamma-ray background.   Line II satisfies the relation between the index of Power-Law Luminosity function \& the width $Log {L_{max}\over L_{min}}$ required by the supernova rate $10^{-4}$ h$^3$ Mpc$^{-3}$ yr$^{-1}$.  To the left of this line, the GRB rate is greater than supernova rate. This line crosses the K-S test favorite region for $Log N-Log P$. This indicates if GRBs are associated with Supernovae, the GRB rate is not a problem for fireball model.  Line III satisfies the relation between the index of Power-Law Luminosity function \& the width $Log {L_{max} \over L_{min}}$ required by the rate $10^{-8}$ h$^3$ Mpc$^{-3} $yr$^{-1} $; This is for a jet with Lorentz factor $\sim$ 100 under the assumption of GRB-SN association. This line also crosses the K-S test favorite region at $L_{min}$  $\sim $ $10^ {49}$ erg/s. This lower limit is higher than most of the
seven SN-GRB pairs explosion energy found by Wang et al. (1998).
\label{fig1}}

\plotone{grblumfig.ps}

\end{document}